\documentclass[a4paper,12pt]{article}
\usepackage{amssymb,amsmath,amsfonts,placeins,pbox,multirow,mathtools}
\usepackage{graphicx,rotate,color,slashed,cite,epstopdf,verbatim,url}
\usepackage{array}
\usepackage[colorlinks=true,
            linkcolor=magenta,
            urlcolor=blue,
            citecolor=blue]{hyperref}
\usepackage[utf8x]{inputenc}

\numberwithin{equation}{section}

\long\def\change#1!!!#2!!!{\color{red}#1 \color{blue}#2 \color{black}}

\def\Eqn#1{Eq.\ (\ref{#1})}
\def\sec#1{Sec.\,\ref{#1}}
\def\reftable#1{Table\ \ref{#1}}

\def\Gsm{{\ensuremath{{\rm SU(3)}_c \times {\rm SU(2)}_L \times {\rm
        U(1)}_Y}}}
\def\rep#1{\ensuremath{\mathit{#1}}}
\def\grp#1{\ensuremath{\{#1\}}}
\def\ssbh#1//#2//{\ensuremath{\xrightarrow [\substack{#2}]
    {\parbox{3cm}{\hfil $\scriptstyle \langle #1 \rangle$ \hfil}}}}

\let\CapTion=\caption
\def\caption#1{\CapTion{\em #1}}

\evensidemargin=\oddsidemargin
\title{\bf SO(10) unification with horizontal symmetry}

\author{\bf Avik Banerjee$^a$\thanks{avik.banerjeesinp@saha.ac.in}, Gautam
  Bhattacharyya$^a$\thanks{gautam.bhattacharyya@saha.ac.in}, Palash
  B. Pal$^b$\thanks{palashbaran.pal@saha.ac.in}
  \\ \medskip \\
  {}$^{a)}$ \small\em Saha Institute of Nuclear
  Physics, HBNI, 1/AF Bidhan Nagar, Kolkata 700064, India \\
  {}$^{b)}$ \small\em Department of Physics, University of Calcutta,
  92 APC Road, Kolkata 700009, India }

\date{}

\begin{document}

\maketitle


\begin{abstract}

We extend the nonsupersymmetric $\rm SO(10)$ grand unification
theories by adding a horizontal symmetry, which connects the three
generations of fermions. Without committing to any specific symmetry
group, we investigate the 1-loop renormalization group evolutions
of the gauge couplings with one and two intermediate breaking
scales. We find that depending on the $\rm SO(10)$ breaking chains,
gauge coupling unification is compatible with only a handful of
choices of representations of the Higgs bosons under the horizontal
symmetry.

\end{abstract}


\bigskip


\section{Introduction}
The quest for the unification of fundamental forces remains to be a
major motivation over the last few decades for investigating theories
beyond the Standard Model.  Although the smallest possible
group, SU(5) \cite{Georgi:1974sy}, has been ruled out, the idea of
unification is very much alive, resting on supersymmetric versions,
and/or bigger unification groups like SO(10) \cite{Georgi:1974my,
  Fritzsch:1974nn,SLANSKY19811,Aulakh:1982sw, Babu:1984mz,
  Deshpande:1992au, Deshpande:1992em, Aulakh:2002zr, Aulakh:2003kg,
  Fukuyama:2004ps, Babu:2005gx, Bertolini:2006pe,
  Bertolini:2009qj,Croon:2019kpe}.  In fact, SO(10) is the smallest
unification group that contains all fermion fields of a single
generation in one irreducible multiplet (irrep).

However, even a group like SO(10) cannot explain why there are three
generations of fermions.  Three copies of the irrep are obviously
needed for three generations, though the number is theoretically
arbitrary.  Therefore, there is a lot of speculation whether an
enhanced symmetry might shed some extra light on the number of fermion
generations \cite{Camargo-Molina:2016bwm,Camargo-Molina:2016yqm,Camargo-Molina:2017kxd,Morais:2020ypd,Morais:2020odg}.

In this paper, we consider the possibility that the three fermion
generations form a multiplet under some horizontal symmetry group $H$
that appears as a direct product with SO(10).  In other words, the
symmetry group of our model is ${\rm SO(10)} \times H$.  We will
examine the unification of the gauge coupling constants of the
Standard Model (SM) into SO(10).  In this pursuit, we do not need to
assume anything specific about the nature of $H$.  For example, we do
not need to know whether $H$ is a global symmetry or a gauge symmetry.
Even more, we do not even need to specify whether it is a discrete
group or a continuous group, or whether the three fermion generations
comprise an irrep under this group.  Multiplets of $H$ will only
contribute in our analysis by providing multiple copies of the Higgs
boson representations, which affect the evolution of the couplings of
SM.  Several earlier attempts have been made to use horizontal
symmetries in explaining the flavor hierarchy in the fermion sector of
the SM \cite{Grimus:2006bb, Bazzocchi:2008sp, Hagedorn:2008bc,
  Grimus:2008tm, Albaid:2009uv, King:2009mk, King:2009tj,
  Dutta:2009bj, Joshipura:2010qx, Patel:2010hr, BhupalDev:2012nm,
  Ferreira:2015jpa, Ivanov:2015xss, Bajc:2016eiw, Bjorkeroth:2017ybg,
  deAnda:2017yeb, King:2017guk}. However, in most of those cases, the
horizontal symmetry is imposed on the fermion mass matrix to generate
the required texture.   Despite these attempts, it is fair to say that 
the flavor problem has not yet been solved, and at the moment there is
nothing close to what may be called the Standard Model of flavor
physics. This necessitates a systematic investigation of all possible
flavor avenues. In this paper, we rely on two basic working
principles, namely, SO(10) unification together with families, and
admission of only renormalizable Yukawa operators.   We will
see that,  without delving into the actual flavor issues, we could
discriminate scenarios and disfavor quite a few of them based on our
working assumptions. This of course narrows down the hunt for the yet
elusive solution to the flavor problem.

The paper is organized as follows. In \sec{s:model}, we outline the
general features of the model, with a short description of SO(10)
unification, followed by how the horizontal symmetry group works.  In
\sec{s:break}, we introduce the various chains of symmetry breaking,
indicating the Higgs boson representation whose vacuum expectation
value (VEV) is responsible for each breaking, and the masses of the
scalar bosons. In \sec{s:results}, we present all of our results for the
symmetry breaking scales.  These results show which chains are
compatible with unification, and what would be the various scales of
symmetry breaking in these chains.  We end with some comments and
outlook in \sec{s:discu}.

\section{$\mathbf{SO(10)}$ with horizontal
  symmetry}\label{s:model} 
In SO(10) grand unified theory (GUT), all left-chiral fermion and
antifermion fields belonging to the same generation constitute a
16-dimensional irrep.  All fermions obtain mass at the level of
electroweak symmetry breaking.  The only exception is the left-chiral
antineutrino field (conjugate of the right-chiral neutrino field),
which is a singlet under the SM gauge group, and can therefore obtain
a Majorana mass at a higher level of symmetry breaking.

The model that we explore in this paper has a symmetry group ${\rm
  SO}(10) \times H$, where $H$ acts between generations.  Our primary
aim in this paper is to examine what multiplicity of various Higgs
boson representations of SO(10) can be accommodated between different
mass thresholds so that their contributions to the renormalization
group (RG) evolution still lead to successful gauge unification. We
draw reference to $H$ only to give credence to this multiplicity of
Higgs bosons. Our conclusions would be based primarily on the number
of appropriately light Higgs states in different stages of RG
evolution, and not so much on what $H$ is and not at all on how
exactly it breaks.

To explain this point, let us consider the couplings of the fermion
multiplets with the scalars, which can be generically written as
$\Psi\Psi\Phi$.  Since the fermions of a single generation transform
like a 16 representation of SO(10), the Higgs bosons coupling to the
fermions must belong to the direct product $16 \times 16$ in order
that the coupling can be a singlet of SO(10).  In SO(10),
\begin{eqnarray}
  16 \times 16 = 10 + 120 + 126 \,,
\end{eqnarray}
of which the first and the last irreps shown on the right side
constitute symmetric combinations, while the 120 is
antisymmetric.  We will consider only the symmetric combinations,
since the 120 is not necessary for conducting the gauge symmetry
breaking of SO(10) down to the electroweak symmetry breaking level.

The three generations of fermions must transform like a
\rep{3}-dimensional representation of the horizontal symmetry group.
As already stated, it is not relevant for our
analysis whether it is reducible or irreducible under $H$.

In writing the Yukawa couplings $\Psi\Psi\Phi$, we have omitted all
indices.  There are the Lorentz indices, the SO(10) indices, as well
as the generation indices.  The combination $\Psi\Psi$ should more
explicitly be written as $\Psi^\top C \Psi$ where $C$ is a matrix,
defined in such a way that $\Psi^\top C$ transforms like
$\overline{\Psi}$ under Lorentz transformation. The combination
$\Psi^\top C \Psi$ must be antisymmetric in the two $\Psi$'s because
the matrix $C$ is always antisymmetric, irrespective of the
representation used for the Dirac matrices \cite{Pal:2007dc}.  We have
chosen to work with symmetric SO(10) combinations.  Thus, according to
the Pauli principle which dictates the overall exchange property to be
antisymmetric with respect to the exchange of the two $\Psi$'s, the
combination $\Psi\Psi$ must be symmetric under the horizontal group.

This requirement does not uniquely determine the dimensionality of the
scalar representations under the horizontal group.  As we said
earlier, we make no assumption regarding whether the three generations
of fermions constitute an irrep of $H$.  All we know, without
specifying the horizontal group, is that the generations form a
\rep{3}-dimensional representation, which can be irreducible or
reducible.  Specification of the horizontal group then dictates
whether that \rep{6}-dimensional representation, obtained by taking
the symmetric product of two \rep{3}-dimensional representations, is
irreducible or reducible.  Accordingly, the representation of $\Phi$
might be irreducible, or might be one of the irreps appearing in the
product $\Psi\Psi$.

It is easy to see that different choices of the group $H$ can lead to
all sorts of dimensions for the irreps.  For example, if the
horizontal symmetry group is SU(3), the fermion generations form a
\rep{3}-dimensional irrep, and the symmetric product is a
\rep{6}-dimensional irrep.  If the symmetry group is SO(3) and the
fermion generations form an irrep, the symmetric product contains a
\rep{5}-dimensional and a \rep{1}-dimensional irrep.  However, in this
case there is also the possibility that the fermion generations do not
form an irrep.  If they transform as a $\rep{2+1}$ dimensional
representation, i.e., one \rep{2}-dimensional irrep and a
\rep{1}-dimensional irrep, then the symmetric product includes
\rep{3}, \rep{2} and \rep{1} dimensional irreps.  In the discrete
group $A_4$, the symmetric product of $\rep{3\times3}$ contains a
\rep{3}-dimensional irrep, plus three different \rep{1}-dimensional
irreps.  If we choose $S_3$ as the horizontal group, which does not
have a \rep{3}-dimensional irrep, the representation of the fermions
can at best be $\rep{2+1}$ dimensional.  The symmetric product of such
a reducible representation with itself contains both $\rep2$ and
$\rep1$.  In short, from symmetry consideration alone, we see anything
from a \rep1-dimensional up to a \rep6-dimensional representation is
allowed for $\Phi$.

If we use a \rep{n}-dimensional representation for $\Phi$, with
$\rep{1\leq n \leq 6}$, then there are \rep{n} copies of the SO(10)
Higgs multiplets of that representation.  These particles participate
in the RG evolution above their mass thresholds, and affect gauge
coupling unification.  The demand of unification imposes what kind of
representations are allowed for $\Phi$.

So far, we have been talking about Higgs multiplets which participate
in Yukawa interaction.  Symmetry breaking of the grand unified group
SO(10) require other Higgs multiplets as well, as we will see in
Section~\ref{s:break}.  All other Higgs bosons, which do not couple to fermions, are assumed to be singlets of the
horizontal group in our discussion.

\section{$\mathbf{SO(10)}$ breaking chains}\label{s:break}
SO(10) might break to the SM gauge group through only one intermediate
group, i.e., the symmetry breaking chain may be of the form
\begin{eqnarray}
  {\rm SO(10)} \stackrel{M_U}{\longrightarrow}  G_1
  \stackrel{M_1}{\longrightarrow}  \Gsm
  \stackrel{M_Z}{\longrightarrow}  {\rm SU(3)}_c \times {\rm 
    U(1)_{em}} \,,
  \label{G1chain}
\end{eqnarray}
where the symbols above the arrows represent the symmetry breaking
scales.  One possibility is that $G_1 = {\rm SU}(5)$.  We ignore this
possibility, since then gauge coupling unification must happen at the
SU(5) level, something that has been ruled out by precision data in
nonsupersymmetric models.  The other possible choices of $G_1$ have
been shown in \reftable{t:1chains}.

From SO(10), the symmetry breaking can also go through two
intermediate scales down to the SM gauge group, i.e., one can consider
symmetry breaking chains like
\begin{eqnarray}
  {\rm SO(10)} \stackrel{M_U}{\longrightarrow} G_2
  \stackrel{M_2}{\longrightarrow}  G_1 \stackrel{M_1}{\longrightarrow}
  \Gsm \stackrel{M_Z}{\longrightarrow}  {\rm SU(3)}_c \times {\rm 
    U(1)_{em}} \,,
  \label{G2G1chain}
\end{eqnarray}
with various choices for the intermediate groups $G_1$ and $G_2$.
These possibilities have been shown in \reftable{t:2chains}.  Breaking
chains with more than two intermediate scales have been considered
in the literature \cite{Bandyopadhyay:2015fka} as well, but we do not
discuss them.

\begin{table}
\caption{Possible chains of SO(10) breaking with one intermediate
  scale.  $G_U$ is the unification group.  In the column SSB1, the
  number above the arrow sign gives the irrep of SO(10) whose VEV is
  responsible for the spontaneous symmetry breaking.  The symbols
  below the arrows represent the multiplets of $G_1$ which contribute
  to the RG evolution as long as $G_1$ is unbroken. Breaking of $G_1$ is given in \Eqn{G1break}.}\label{t:1chains}
$$
\begin{array}{cccc}
  \hline\hline 
  \mbox{Chain} & G_U & \mbox{SSB1} & G_1 \\
  \hline
  1 & \mbox{SO(10)} & \ssbh 210 // (2,2,1)_{10}
    \\ (1,3,10)_{126} / (1,2,4)_{16} // &
  \{2_L2_R4_C\} \\
  \hline
  2 & \mbox{SO(10)} & \ssbh 54 // (2,2,1)_{10}
    \\ (1,3,10)_{126} / (1,2,4)_{16}
    \\ (3,1,10)_{126} / (2,1,4)_{16} //
  & \{2_L2_R4_CP\}\\
  \hline
  3 & \mbox{SO(10)} & \ssbh 45 // (2,2,0,1)_{10}
    \\ (1,3,1,1)_{126} / (1,2,\frac12,1)_{16} // & \{2_L2_R1_X3_c\}\\
  \hline
  4 & \mbox{SO(10)} & \ssbh 210 // (2,2,0,1)_{10}
    \\ (1,3,1,1)_{126} / (1,2,\frac12,1)_{16}
    \\ (3,1,1,1)_{126} / (2,1,-\frac12,1)_{16} //
  & \{2_L2_R1_X3_cP\}\\
  \hline
  5 & \mbox{SO(10)} & \ssbh 45 // (2,\frac12,1)_{10} \\
    (1,1,10)_{126} / (1,-\frac12,4)_{16} // & \{2_L1_R4_C\}\\
  \hline
  6 & \mbox{SO(10)} & \ssbh 210 // (2,\frac12,0,1)_{10} \\
    (1,1,-1,1)_{126} / (1,-\frac12,\frac12,1)_{16} // &
  \{2_L1_R1_X3_c\}\\
  \hline\hline
\end{array}
$$

\end{table}

\begin{table}
    \caption{Possible chains of SO(10) breaking with two intermediate
      scales.}\label{t:2chains}

    \bigskip

    \begin{tabular}{cccccc}
      \hline\hline
    \mbox{Chain} & $G_U$ & \mbox{SSB2} & $G_2$ & \mbox{SSB1} &
  $G_1$ \\
  \hline

    I & SO(10) & \ssbh 210 //
    (2,2,1)_{10} \\ (1,3,10)_{126} / (1,2,4)_{16} \\ (1,1,15)_{45}
    // & \grp{2_L2_R4_C} & \ssbh 45 //
    (2,2,0,1)_{10} \\ (1,3,-1,1)_{126} / (1,2,\frac12,1)_{16}
    // & \grp{2_L2_R1_X3_c} \\*
    \hline
    
    II & SO(10) & \ssbh 54 //
    (2,2,1)_{10} \\ (1,3,10)_{126} / (1,2,4)_{16} \\ (3,1,10)_{126} /
    (2,1,4)_{16} \\ (1,1,15)_{210}
    // & \grp{2_L2_R4_CP} & \ssbh 210 //
    (2,2,0,1)_{10} \\ (1,3,-1,1)_{126} / (1,2,\frac12,1)_{16} \\
    (3,1,-1,1)_{126} / (2,1,-\frac12,1)_{16}
    // &
    \grp{2_L2_R1_X3_cP}  \\*
    \hline
    
    III & SO(10) & \ssbh 54 //
    (2,2,1)_{10} \\ (1,3,10)_{126} / (1,2,4)_{16} \\ (3,1,10)_{126} /
    (2,1,4)_{16} \\ (1,1,15)_{45}
    // & \grp{2_L2_R4_CP} & \ssbh 45 //
    (2,2,0,1)_{10} \\ (1,3,-1,1)_{126} / (1,2,\frac12,1)_{16}
    // &
    \grp{2_L2_R1_X3_c}  \\*
    \hline

    IV & SO(10) & \ssbh 210 //
    (2,2,0,1)_{10} \\ (1,3,-1,1)_{126} / (1,2,\frac12,1)_{16} \\
    (3,1,1,1)_{126} / (2,1,-\frac12,1)_{16} \\ (1,1,0,1)_{45}
    //  & \grp{2_L2_R1_X3_cP}  & \ssbh 45 //
    (2,2,0,1)_{10} \\ (1,3,-1,1)_{126} / (1,2,\frac12,1)_{16}
    // & \grp{2_L2_R1_X3_c}  \\*
    \hline

    V & SO(10) & \ssbh 210 //
    (2,2,1)_{10} \\ (1,3,10)_{126} / (1,2,4)_{16} \\ (1,3,1)_{45}
    //  & \grp{2_L2_R4_C}  & \ssbh 45 //
    (2,\frac12,1)_{10} \\ (1,1,10)_{126} / (1,-\frac12,4)_{16}
    // &
    \grp{2_L1_R4_C}  \\*
    \hline

    VI & SO(10) & \ssbh 54 //
    (2,2,1)_{10} \\ (1,3,10)_{126} / (1,2,4)_{16}
     \\ (3,1,10)_{126} / (2,1,4)_{16} \\ (1,3,1)_{45} + (3,1,1)_{45}
    //  & \grp{2_L2_R4_CP}  & \ssbh 45 //
    (2,\frac12,1)_{10} \\ (1,1,10)_{126} / (1,-\frac12,4)_{16}
    // &
    \grp{2_L1_R4_C}  \\*
    \hline

    VII & SO(10) & \ssbh 54 //
    (2,2,1)_{10} \\ (1,3,10)_{126} / (1,2,4)_{16}
    \\ (3,1,10)_{126} / (2,1,4)_{16}
    \\ (1,1,1)_{210}
    //  & \grp{2_L2_R4_CP} & \ssbh 210 //
    (2,2,1)_{10} \\ (1,3,10)_{126} / (1,2,4)_{16}
    // &
    \grp{2_L2_R4_C}  \\*
    \hline

    VIII & SO(10) & \ssbh 45 //
    (2,2,0,1)_{10} \\ (1,3,-1,1)_{126} / (1,2,\frac12,1)_{16}
    \\ (1,3,0,1)_{45} 
    //  & \grp{2_L2_R1_X3_c}  & \ssbh 45 //
    (2,\frac12,0,1)_{10} \\ (1,1,-1,1)_{126} / (1,-\frac12,\frac12,1)_{16}
    // &
    \grp{2_L1_R1_X3_c}  \\*
    \hline

    IX & SO(10) & \ssbh 210 //
    (2,2,0,1)_{10} \\ (1,3,-1,1)_{126} / (1,2,\frac12,1)_{16}
    \\ (3,1,1,1)_{126} / (2,1,-\frac12,1)_{16} \\
    (1,3,0,1)_{45} + (3,1,0,1)_{45}
    //  & \grp{2_L2_R1_X3_cP}  & \ssbh 45 //
    (2,\frac12,0,1)_{10} \\ (1,1,-1,1)_{126} / (1,-\frac12,\frac12,1)_{16}
        // &
    \grp{2_L1_R1_X3_c} \\*
    \hline

    X & SO(10) & \ssbh 210 //
    (2,2,1)_{10} \\ (1,3,10)_{126} / (1,2,4)_{16} \\ (1,3,15)_{210}
    //  & \grp{2_L2_R4_C}  & \ssbh 210 //
    (2,\frac12,0,1)_{10} \\ (1,1,-1,1)_{126} / (1,-\frac12,\frac12,1)_{16}
    // &
    \grp{2_L1_R1_X3_c}  \\*
    \hline

    XI & SO(10) & \ssbh 54 //
    (2,2,1)_{10} \\ (1,3,10)_{126} / (1,2,4)_{16}
    \\ (3,110)_{126} / (2,1,4)_{16}
    \\ (1,3,15)_{210}  + (3,1,15)_{210}
    //  & \grp{2_L2_R4_CP}  & \ssbh 210 //
    (2,\frac12,0,1)_{10} \\ (1,1,-1,1)_{126} / (1,-\frac12,\frac12,1)_{16}
    // &
    \grp{2_L1_R1_X3_c}  \\*
    \hline

    XII & SO(10) & \ssbh 45 //
    (2,\frac12,1)_{10} \\ (1,1,10)_{126} / (1,-\frac12,4)_{16}
    \\ (1,0,15)_{45} 
    //  & \grp{2_L1_R4_C}  & \ssbh 45 //
    (2,\frac12,0,1)_{10} \\ (1,1,-1,1)_{126} / (1,-\frac12,\frac12,1)_{16}
    // &
    \grp{2_L1_R1_X3_c}  \\*
    
\hline
\hline
    \end{tabular}
  
\end{table}

In writing \reftable{t:1chains} and \reftable{t:2chains} as well as in
subsequent discussions, we have adopted some shorthand notation used
in many earlier articles on SO(10) breaking \cite{Deshpande:1992au,
  Deshpande:1992em,Bertolini:2009qj}.  For the sake of completeness,
we briefly explain the notations here.
\begin{itemize}

\item For example, let us look at the first chain given in
  \reftable{t:1chains}.  It shows that the intermediate gauge group is
  $2_L2_R4_C$.  This stands for ${\rm SU(2)}_L \times {\rm SU(2)}_R
  \times {\rm SU(4)}_C$, where the SU(4) factor stands for the
  Pati-Salam group \cite{Pati:1974yy} which treats leptons as a fourth
  ``color'', ${\rm SU(2)}_L$ belongs to the SM electroweak gauge
  group, and the other one is a right-handed SU(2) under which the
  right-chiral quark and lepton fields transform separately as
  doublets.  Similarly, presence of the symbol $3_c$ would imply the
  SU(3) gauge group of QCD.  The direct product factor $P$, which
  appears in some chains, is a discrete symmetry between the two SU(2)
  factors, which ensures that the coupling constants of the two
  SU(2)'s are equal.

\item As for $1_X$ which is a U(1) subgroup, the quantum numbers shown
  in \reftable{t:1chains} and \reftable{t:2chains} equal to $(B-L)/2$.
  With this definition, the normalization of $X$ does not agree with
  that of the non-Abelian factors, so the quantum numbers will have to
  be multiplied by a factor of $\sqrt{3/2}$ in order to be used in RG
  equations. For the $1_R$ subgroup however, we put the eigenvalues of
  the corresponding generator of the $2_R$ subgroup, which have the
  proper normalization.

\item In \reftable{t:1chains}, the column bearing the heading `SSB1'
  contains information about symmetry breaking and masses of the Higgs
  bosons.  The number above the arrow gives the SO(10) representation
  which has a neutral component whose VEV can perform the desired
  symmetry breaking.  Of course these multiplets have VEVs at the
  unification scale, and therefore all their components are expected
  to have masses at the unifications scale, meaning that they do not
  affect the RG equations.  But, in the regime below the SO(10)
  breaking scale where the group $G_1$ is the unbroken gauge group,
  the RG equations contain contributions from some Higgs boson
  submultiplets which contain VEVs that affect one of the \emph{lower}
  stages of symmetry breaking.  They have been shown below the arrow,
  indicating their transformations under the unbroken group at that
  stage, and marked with a subscript that tells us which SO(10)
  representation contains them.  The rationale for choosing the masses
  of the Higgs bosons will be described in Section~\ref{s:results}.

\item We have not shown in \reftable{t:1chains} which multiplet of
  SO(10) breaks the intermediated symmetry down to the SM symmetry
  group, and further trigger the electroweak symmetry breaking.  This
  is because this part is the same for all chains, and is given by
\begin{eqnarray}
  G_1 \ssbh 126 / 16 // (2,\frac12,1)_{10}
  // 2_L1_Y3_c \ssbh 10 // // 1_Q3_c \,.
  \label{G1break}
\end{eqnarray}
   As seen
   here, we have considered two alternatives for breaking $G_1$ to the
   SM gauge group, separated by a slash in \Eqn{G1break}.  One is by
   the 126-dimensional representation of SO(10), which has Yukawa
   coupling with fermions.  The other is by using the 16-dimensional
   irrep, which does not have Yukawa coupling at the tree-level, but
   can contribute to fermion masses through loops
   \cite{Witten:1979nr}. In subsequent tables and discussions,
   for the first alternative, the chain is specified by adding the
   letter `a', whereas the letter `b' is added for the second
   alternative.

\item The same notations are used for chains with two intermediate
  scales.  The only difference is that, in this case, one needs to
  specify the Higgs boson multiplets which perform the intermediate
  scale symmetry breaking, $G_2 \to G_1$.  Thus, the column with the
  heading `SSB2' contains information about the Higgs multiplet that
  performs the breaking ${\rm SO(10)} \to G_2$, and the submultiplets
  of $G_2$ which we consider for the RG equations above the
  $G_2$-breaking scale.  Similarly, the column marked `SSB1' contains
  information about the multiplet of SO(10) that is responsible for
  the breaking $G_2 \to G_1$, and the submultiplets of $G_1$ which are
  assumed to be light above the $G_1$-breaking scale.

\end{itemize}

\section{Results}\label{s:results}
\subsection{Outline of the strategy}
The 1-loop RG evolution of the gauge coupling $g$ for an ${\rm SU}(N)$
factor above the weak scale is governed by the equation
\begin{eqnarray}
{d\omega \over d \ln \mu} = {1 \over 2\pi} \Big( 
\frac{11}3 N - 4 - {T(S) \over 6} \Big) \,,
\label{rgeq}
\end{eqnarray}
where 
\begin{eqnarray}
\omega = {4\pi \over g^2} \,, 
\label{omega}
\end{eqnarray}
and $T(S)$ is the scalar contribution.  For a U(1) factor in the
symmetry group, one should take $N=0$ in \Eqn{rgeq}.  Note that, in
writing the fermion contribution in \Eqn{rgeq}, we assumed for
simplicity that symmetry breaking scales are heavier than the masses
of all fermions which obtain masses at that stage of breaking.  In
fact, we consider the right-handed neutrinos to have Majorana masses
which are smaller than $M_1$, the scale at which all symmetries
specific to the right-handed fermions break down in the chain of
symmetry breaking from SO(10) , i.e. below which the SM group appears.

For the scalar contribution, we need to have some idea of the Higgs
boson masses.  As is usual practice in this kind of analysis, we use
the extended survival hypothesis \cite{delAguila:1980qag,
  Mohapatra:1982aq} to estimate the masses.  This means that, at any
stage of symmetry breaking, the entire submultiplet containing the VEV
obtains mass at that scale, and any particle not controlled by this
rule obtains mass at the unification scale.  We have further extended
this hypothesis across generations, i.e. whatever is the mass
hierarchy generated due to SO(10) breaking for one family is
replicated for other (at least one more) families so as to generate
Higgs multiplicities.
We have already listed, in \reftable{t:1chains} and
\reftable{t:2chains}, the Higgs boson submultiplets which contribute
to $T(S)$ in all different regimes.  The overall contributions to
$T(S)$ for all intermediate symmetry groups have been listed in
earlier literature \cite{Deshpande:1992au, Deshpande:1992em} where
only one 10-dimensional and one 126-dimensional SO(10) multiplets of
Higgs boson were considered.  For the present purpose, all we need is
to multiply the contributions of 10 and 126 irreps by the appropriate
number of multiplets. 

So we start with the values of the gauge coupling constants of the SM
gauge group at the $Z$-scale \cite{Tanabashi:2018oca}:
\begin{subequations}
  \label{omegaexp}
\begin{eqnarray}
\omega_{\rm 1_Y}(M_Z)&=& 59.042\pm 0.003, \\
\omega_{\rm 2_L}(M_Z)&=& 29.596\pm 0.005, \\
\omega_{\rm 3_c}(M_Z)&=& 8.47\pm 0.02\,.
\end{eqnarray}
\end{subequations}
As has already been said, the evolution depends on the chain of
symmetry breaking, for which we use the nomenclature used by earlier
authors and repeated here in \reftable{t:1chains} and
\reftable{t:2chains}.  In addition, the evolution depends on the
numbers of 10 and 126 irreps of Higgs bosons used, which we will
denote by $r_{10}$ and $r_{126}$ respectively.  For each chain, and
each choice of $r_{10}$ and $r_{126}$, we employ the following checks
to determine whether a given chain is allowed.
\begin{itemize}
  \item There must be a solution for all gauge couplings meeting at a
    scale.

  \item When a mathematical solution is obtained, it must be
    physically meaningful.  For example, in \Eqn{G2G1chain}, if SO(10)
    breaks at $M_U$, whereas $G_2$ and $G_1$ break at the scales $M_2$
    and $M_1$ respectively, one must have 
    \begin{eqnarray}
      n_U > n_2 > n_1 > n_Z \,,
    \end{eqnarray}
    where the $n_i$'s represent a logarithmic notation that we use for
    the various energy scales:
    \begin{eqnarray}
      n_i = \log_{10} \left( M_i \over 1\;\mbox{GeV} \right) \,.
    \end{eqnarray}
    
 \item   
 	Further, one must have $M_U \lesssim 10^{18}$\,GeV because
        gravity effects become strong at higher scales, and the
        analysis that ignores gravity makes no sense.

  \item 
    The unification scale must be consistent with proton decay
    bound. The lifetime of proton in terms of the GUT scale and
    couplings can be written as
    \begin{eqnarray}
    	\tau_p\simeq \frac{\omega_U^2M_U^4}{m_p^5}\, ,
    \end{eqnarray}
    where $m_p$ denotes proton mass.  Present limit on the proton
    lifetime is \cite{Tanabashi:2018oca}
    \begin{eqnarray}
      \tau_p(p\to e^+\pi^0) > 1.6\times10^{34}~{\rm yr}.
      \label{pdkexp}
    \end{eqnarray}
    However, taking cue from \cite{Bertolini:2009qj}, we have made an
    order of magnitude estimate to accommodate the higher-loop effects
    in the proton decay bound. Our estimate shows that proton lifetime
    as calculated using 1-loop results for the GUT parameters can be
    enhanced by two orders of magnitude if 2-loop effects are
    incorporated.  In view of this conservative estimation, we also
    allow those chains which satisfy
    \begin{eqnarray}
      \tau_p(p\to e^+\pi^0) > 1.6\times10^{32}~{\rm yr}.
      \label{pdkexprel}
    \end{eqnarray}
    We clearly mark which chains satisfy \Eqn{pdkexp} and which ones
    satisfy only this relaxed bound, \Eqn{pdkexprel}.

  \item All couplings are consistent with the perturbative limit
    \begin{eqnarray}
      \omega > {1 \over 4\pi} \,.
    \end{eqnarray}
\end{itemize}
If any one of these conditions is not satisfied for a chain, the chain
is ruled out, and its details are not given in the tables.

In our calculations, whenever it is relevant, we take kinetic mixing
of U(1) gauge factors into account, something that was not done in the
early papers on the subject but was incorporated in the later papers
\cite{Holdom:1985ag, delAguila:1995rb, Luo:2002iq, Bertolini:2009qj}.
Our results for the allowed chains, and the allowed values of various
scales of symmetry breaking, are presented in the rest of this
section.  For the sake of convenience, we divide the discussion into
two parts: one in which there is only one intermediate scale, and the
other in which there are two such scales.

\begin{table}[p]
\centering

\def\pdb#1{\ifnum #1=0 $\downarrow$
  \else \phantom{e} \fi}

\caption{Results for chains with a-type symmetry breaking with one
  intermediate scale: intermediate scale ($n_1$), unification scale
  ($n_U$) and unification coupling ($\omega_U$) have been shown for
  each case where we find an acceptable solution.  Nomenclatures for
  the chains can be read off from \reftable{t:1chains}. A down arrow to
  the right of the value of $n_U$ indicates that the given combination
  of unification scale and coupling does not satisfy the proton
  lifetime bound in \Eqn{pdkexp}, but allowed by the conservative bound in
  \Eqn{pdkexprel}.}\label{t:a1results}

\bigskip

\begin{tabular}{ccrrrr}
  \hline\hline
  \multirow2*{($r_{10}$, $r_{126}$)}
  & \multirow2*{Quantity} & \multicolumn{4}{c}{Breaking
    chains} \\  

  \cline{3-6}

  & & \multicolumn1c{1a\pdb1} & \multicolumn1c{2a\pdb1}
  & \multicolumn1c{3a\pdb1} & \multicolumn1c{4a\pdb1}
  \\ 

  \hline
  
  \multirow3*{\rep{(1,1)}} 
  & $n_1$      & 11.75\pdb1 & 13.71\pdb1 &  9.02\pdb1 & 10.11\pdb1 \\
  & $n_U$      & 16.06\pdb1 & 15.22\pdb0 & 16.66\pdb1 & 15.77\pdb1 \\
  & $\omega_U$ & 45.70\pdb1 & 41.20\pdb1 & 46.19\pdb1 & 43.89\pdb1 \\

  \hline
  
  \multirow3*{\rep{(1,2)}} 
  & $n_1$     && &  6.16\pdb1 \\
  & $n_U$     && & 16.54\pdb1 \\
  & $\omega_U$&& & 45.89\pdb1 \\

  \hline
  
  \multirow3*{\rep{(2,1)}} 
  & $n_1$      & 12.19\pdb1 & & 10.24\pdb1 & 10.95\pdb1 \\
  & $n_U$      & 15.53\pdb0 & & 15.90\pdb1 & 15.29\pdb0 \\
  & $\omega_U$ & 44.11\pdb1 & & 44.23\pdb1 & 42.66\pdb1 \\

  \hline
  
  \multirow3*{\rep{(2,2)}} 
  & $n_1$      && &  8.28\pdb1 \\
  & $n_U$      && & 15.75\pdb1 \\
  & $\omega_U$ && & 43.84\pdb1 \\

  \hline
  
  \multirow3*{\rep{(2,3)}} 
  & $n_1$      && &  4.49\pdb1 \\
  & $n_U$      && & 15.46\pdb0 \\
  & $\omega_U$ && & 43.09\pdb1 \\

  \hline
  
  \multirow3*{\rep{(3,1)}} 
  & $n_1$      && & 11.19\pdb1 \\
  & $n_U$      && & 15.30\pdb0 \\
  & $\omega_U$ && & 42.69\pdb1 \\

\hline\hline

\end{tabular}
\end{table}

\begin{table}[p]
\centering

\def\pdb#1{\ifnum #1=0 $\downarrow$
  \else \phantom{e} \fi}

\caption{Results for chains with b-type symmetry breaking with one
  intermediate scale.  Notations are same as in
  \reftable{t:a1results}.}\label{t:b1results}

\bigskip

\begin{tabular}{ccrrrr}
  \hline\hline
  \multirow2*{($r_{10}$)}
  & \multirow2*{Quantity} & \multicolumn{4}{c}{Breaking
    chains} \\  

  \cline{3-6}

  & & \multicolumn1c{1b} & \multicolumn1c{2b}
  & \multicolumn1c{3b} & \multicolumn1c{4b}
  \\ 

  \hline
  
  \multirow3*{\rep{(1)}} 
  & $n_1$       & 13.63\pdb1 & 13.71\pdb1 & 10.41\pdb1 & 10.67\pdb1 \\
  & $n_U$       & 15.39\pdb0 & 15.35\pdb0 & 16.72\pdb1 & 16.49\pdb1 \\
  & $\omega_U$  & 45.07\pdb1 & 44.64\pdb1 & 46.34\pdb1 & 45.75\pdb1 \\

  \hline
  
  \multirow3*{\rep{(2)}} 
  & $n_1$     && & 11.23\pdb1 & 11.40\pdb1 \\
  & $n_U$     && & 15.97\pdb1 & 15.82\pdb1 \\
  & $\omega_U$&& & 44.42\pdb1 & 44.02\pdb1 \\

  \hline
  
  \multirow3*{\rep{(3)}} 
  & $n_1$      && & 11.89\pdb1 & 12.00\pdb1 \\
  & $n_U$      && & 15.37\pdb0 & 15.27\pdb0 \\
  & $\omega_U$ && & 42.89\pdb1 & 42.62\pdb1 \\

\hline\hline

\end{tabular}
\end{table}

\subsection{One intermediate scale}
If there is only one intermediate scale, the possibilities of the
intermediate gauge group have been shown in \reftable{t:1chains}.  The
results obtained from the 1-loop RG equations have been summarized in
two tables, where we show the intermediate scale, the unification
scale and the unification coupling for all chains which admit an
acceptable solution.  In \reftable{t:a1results}, we consider the SM
symmetry group to appear through the breaking using 126 multiplet of
Higgs bosons, whereas in \reftable{t:b1results}, we consider the
possibility that the said intermediate symmetry breaking occurs
through a 16 multiplet of Higgs bosons.  If any chain does not appear
in these tables, or there is no entry for the scales corresponding to
any chain, it means that there is no solution consistent with the
criteria outlined earlier.

For all chains, the case with $r_{10}=r_{126}=\rep1$ takes us to the
limiting scenario with no horizontal symmetry.  The differences of our
results with that obtained earlier \cite{Deshpande:1992au} using
126-irrep are because of two reasons.  First, we use updated inputs
for the couplings at the weak scale, given in \Eqn{omegaexp} and for
the constraint on proton lifetime. Second, while we have performed
only a 1-loop calculation here, 2-loop results are available in the
literature \cite{Deshpande:1992au}.  However, note that we keep some
room for the 2-loop effects using some simple estimates as mentioned
earlier.

We now highlight the new results found in the present paper.  First,
in a-type chains, we analyze the effects of the values of $r_{10}$ and
$r_{126}$ larger than \rep1, which are necessary for discussing
horizontal symmetry.  Second, cases for b-type symmetry breaking
\cite{Rajpoot:1980xy, Fukugita:2003en}, which use a 16 instead of a
126 for breaking down to the level of the SM symmetry group, have been
analyzed here with $r_{10}> \rep{1}$.

Note that we obtain no solution at all for chains 5 and 6 in
\reftable{t:1chains}.  This means that, if there is only one
intermediate stage between the GUT group and the SM group, then that
intermediate stage must contain the full ${\rm SU(2)}_R$ symmetry.

If we look at models where at least one of $r_{10}$ and $r_{126}$ is
bigger than \rep1, we see that chain 2 is also ruled out, and chain 1a
is allowed only when $r_{10}=\rep2$.  This means that the full ${\rm
  SU(4)}_C$ symmetry at the intermediate stage is disfavored, or
equivalently that the ${\rm SU(3)}_c$ group of QCD should appear right
at the grand unified symmetry breaking.

It is also seen that there is no solution in any symmetry breaking
chain when either $r_{10}$ and $r_{126}$ is bigger than \rep3.  The
intermediate group $2_L2_R1_X3_c$ seems to be most suitable, in the
sense that it is most flexible with the numbers of the Higgs boson
multiplets.  Its companion version with the extra parity symmetry does
equally well for b-type chains, but not so well with the a-type
chains. 


\def\pdk#1#2{%
\ifnum #1=1 \ifnum #2=1 --\fi\fi
\ifnum #1=0 \ifnum #2=1 \raise2pt\hbox{\tiny$\leftarrow$} \fi\fi
\ifnum #1=1 \ifnum #2=0 \raise2pt\hbox{\tiny$\rightarrow$} \fi\fi
\ifnum #1=0 \ifnum #2=0 \raise2pt\hbox{\tiny$\leftrightarrow$} \fi\fi
}

\def\fake{\phantom{00.00}}

\begin{table}
\caption{ The range of allowed values for a-type symmetry breaking
with two intermediate scales. Nomenclatures for the chains
can be read off from \reftable{t:2chains}.  In case of $n_U$, an arrow on either side of
the separator points to the combination of unification scale and coupling that does not
satisfy the proton lifetime bound in \Eqn{pdkexp}, but allowed
by the conservative bound in \Eqn{pdkexprel}.}\label{t:a2results}

\bigskip

\footnotesize
\begin{center}
\begin{tabular}{ccc@{\hspace{1pt}}c@{\hspace{1pt}}cc@{\hspace{1pt}}c@{\hspace{1pt}}cc@{\hspace{1pt}}c@{\hspace{1pt}}cc@{\hspace{1pt}}c@{\hspace{1pt}}cc@{\hspace{1pt}}c@{\hspace{1pt}}c}

  \hline\hline
  \multicolumn{17}c{Chains Ia to Va} \\ 
  \hline\hline

  $(r_{10},r_{126})$
  & Quantity & \multicolumn3c{Ia} &
  \multicolumn3c{IIa} & \multicolumn3c{IIIa}
  & \multicolumn3c{IVa} & \multicolumn3c{Va}  \\
  
  \hline

  \multirow4*{\rep{(1,1)}}
  & $n_1$     &  9.02&  --  & 11.17 & 10.12&  --  & 13.70  &  9.02&  --  & 13.70  &  9.02&  --  & 10.11  & 11.36&  --  & 11.58  \\ 
  & $n_2$     & 16.66&  --  & 11.19 & 15.77&  --  & 13.71  & 16.66&  --  & 13.72  & 16.65&  --  & 10.15  & 13.63&  --  & 11.66  \\ 
  & $n_U$     & 16.66&\pdk11& 16.75 & 15.77&\pdk10& 15.35  & 16.66&\pdk10& 15.36  & 16.66&\pdk11& 15.77  & 15.10&\pdk01& 16.00  \\ 
  & $\omega_U$& 46.19&  --  & 46.42 & 43.89&  --  & 41.03  & 46.19&  --  & 41.03  & 46.19&  --  & 43.91  & 44.75&  --  & 45.62  \\

  \hline

  \multirow4*{\rep{(1,2)}}
  & $n_1$      &  2.00&  --  &  6.15 & 12.66&  --  & 13.71 &  6.16&  --  & 13.71 &  6.16&  --  &  9.17 \\
  & $n_2$      & 15.94&  --  & 16.54 & 14.02&  --  & 13.71 & 16.54&  --  & 13.71 & 16.54&  --  & 10.49 \\
  & $n_U$      & 17.67&\pdk11& 16.54 & 15.14&\pdk00& 15.22 & 16.54&\pdk10& 15.22 & 16.54&\pdk10& 15.13 \\
  & $\omega_U$ & 46.87&  --  & 45.89 & 38.59&  --  & 37.51 & 45.88&  --  & 37.51 & 45.88&  --  & 42.25 \\

  \hline

  \multirow4*{\rep{(1,3)}}
  & $n_1$      &&& &&&  &  2.00&  --  & 13.08 &  2.00&  --  &  5.06\\
  & $n_2$      &&& &&&  & 15.93&  --  & 13.83 & 14.78&  --  & 12.38\\
  & $n_U$      &&& &&&  & 16.12&\pdk10& 15.16 & 15.84&\pdk10& 15.12\\
  & $\omega_U$ &&& &&&  & 43.72&  --  & 35.01 & 44.08&  --  & 42.24\\

  \hline

  \multirow4*{\rep{(2,1)}}
  & $n_1$      & 10.24&  --  & 11.79 & 10.95&  --  & 12.72 & 10.24&  --  & 13.34 & 10.24&  --  & 10.95 & 11.91&  --  & 12.07 \\
  & $n_2$      & 15.89&  --  & 11.79 & 15.29&  --  & 14.28 & 15.90&  --  & 13.94 & 15.87&  --  & 10.98 & 12.99&  --  & 12.09 \\
  & $n_U$      & 15.90&\pdk11& 16.02 & 15.29&\pdk00& 15.13 & 15.90&\pdk10& 15.13 & 15.89&\pdk10& 15.29 & 15.12&\pdk00& 15.50 \\
  & $\omega_U$ & 44.23&  --  & 44.53 & 42.66&  --  & 41.32 & 44.22&  --  & 40.96 & 44.22&  --  & 42.67 & 43.80&  --  & 44.07 \\

  \hline

  \multirow4*{\rep{(2,2)}}
  & $n_1$      &  2.00&  --  &  8.28 &&& &  2.00&  --  &  8.28 &  8.28&  --  &  9.55 \\
  & $n_2$      & 14.34&  --  & 15.75 &&& & 14.34&  --  & 15.75 & 15.74&  --  & 12.78 \\
  & $n_U$      & 17.35&\pdk11& 15.75 &&& & 17.35&\pdk11& 15.75 & 15.75&\pdk10& 15.12 \\
  & $\omega_U$ & 44.64&  --  & 43.84 &&& & 44.64&  --  & 43.84 & 43.84&  --  & 42.24 \\

  \hline

  \multirow4*{\rep{(2,3)}}
  & $n_1$      &  2.00&  --  &  4.49 &&& &  4.49&  --  &  9.28 &  4.49&  --  &  5.84 \\
  & $n_2$      & 15.39&  --  & 15.46 &&& & 15.45&  --  & 14.55 & 15.45&  --  & 14.18 \\
  & $n_U$      & 15.72&\pdk10& 15.46 &&& & 15.45&\pdk00& 15.14 & 15.45&\pdk00& 15.12 \\
  & $\omega_U$ & 43.05&  --  & 43.09 &&& & 43.09&  --  & 39.04 & 43.09&  --  & 42.24 \\

  \hline

  \multirow4*{\rep{(3,1)}}
  & $n_1$      & 11.19&  --  & 12.29 &&& & 11.19&  --  & 11.99 & 11.19&  --  & 11.38 \\
  & $n_2$      & 15.30&  --  & 12.30 &&& & 15.30&  --  & 14.80 & 15.30&  --  & 13.75 \\
  & $n_U$      & 15.30&\pdk00& 15.42 &&& & 15.30&\pdk00& 15.13 & 15.30&\pdk00& 15.12 \\
  & $\omega_U$ & 42.69&  --  & 43.00 &&& & 42.69&  --  & 41.88 & 42.69&  --  & 42.25 \\

  \hline

  \multirow4*{\rep{(3,2)}}
  & $n_1$      &  2.00&  --  &  9.85 &&& &  9.86&  --  & 10.11 &  9.86&  --  &  9.92 \\
  & $n_2$      & 12.80&  --  & 15.16 &&& & 15.16&  --  & 15.06 & 15.15&  --  & 14.99 \\
  & $n_U$      & 17.04&\pdk10& 15.16 &&& & 15.16&\pdk00& 15.12 & 15.16&\pdk00& 15.13 \\
  & $\omega_U$ & 42.50&  --  & 42.33 &&& & 42.32&  --  & 42.04 & 42.32&  --  & 42.25 \\

  \hline

  \multirow4*{\rep{(3,3)}}
  & $n_1$      &  2.00&  --  &  4.95 \\
  & $n_2$      & 14.56&  --  & 14.74 \\
  & $n_U$      & 15.40&\pdk00& 15.13 \\
  & $\omega_U$ & 41.10&  --  & 41.40 \\

  \hline

  \multirow4*{\rep{(4,2)}}
  & $n_1$      &  2.00&  --  &  9.17 \\
  & $n_2$      & 11.32&  --  & 13.99 \\
  & $n_U$      & 16.75&\pdk10& 15.13 \\
  & $\omega_U$ & 40.44&  --  & 41.01 \\

\hline\hline

\multicolumn{15}r{(Table continued to page \pageref{p:a2p2})}
\label{p:a2p1}

\end{tabular}
\end{center}

\end{table}

\begin{table}[p]
  \label{p:a2p2}

\bigskip

\footnotesize
\begin{center}
\begin{tabular}{ccc@{\hspace{1pt}}c@{\hspace{1pt}}cc@{\hspace{1pt}}c@{\hspace{1pt}}cc@{\hspace{1pt}}c@{\hspace{1pt}}cc@{\hspace{1pt}}c@{\hspace{1pt}}cc@{\hspace{1pt}}c@{\hspace{1pt}}c}

  \multicolumn7l{(\reftable{t:a2results}: continued from 
    page \pageref{p:a2p1})}  \\ 

&& \fake && \fake & \fake && \fake & \fake && \fake & \fake && \fake
& \fake && \fake \\

  \hline\hline
  \multicolumn{17}c{Chains Ia to Va} \\ 
  \hline\hline

  $(r_{10},r_{126})$
  & Quantity & \multicolumn3c{Ia} &
  \multicolumn3c{IIa} & \multicolumn3c{IIIa}
  & \multicolumn3c{IVa} & \multicolumn3c{Va}  \\
  
  \hline

  \multirow4*{\rep{(5,2)}}
  & $n_1$      &  2.00&  --  &  8.26 \\
  & $n_2$      &  9.90&  --  & 12.67 \\
  & $n_U$      & 16.46&\pdk10& 15.14 \\
  & $\omega_U$ & 38.47&  --  & 39.57 \\

  \hline

  \multirow4*{\rep{(6,2)}}
  & $n_1$      &  2.00&  --  &  7.26 \\
  & $n_2$      &  8.54&  --  & 11.22 \\
  & $n_U$      & 16.19&\pdk10& 15.15 \\
  & $\omega_U$ & 36.56&  --  & 37.98 \\

\hline\hline

\\  \\ 

  \hline\hline
  \multicolumn{17}c{Chains VIa to Xa} \\ 
  \hline\hline
  $(r_{10},r_{126})$
  & Quantity & \multicolumn3c{VIa} &
  \multicolumn3c{VIIa} & \multicolumn3c{VIIIa}
  & \multicolumn3c{IXa} & \multicolumn3c{Xa}  \\
  
  \hline

  \hline

  \multirow4*{\rep{(1,1)}}
  & $n_1$      & 13.55&  --  & 13.71 & 11.75&  --  & 13.71 &  2.00&  --  &  8.57 &  2.00&  --  & 10.58 \\
  & $n_2$      & 13.77&  --  & 13.71 & 16.06&  --  & 13.71 &  7.89&  --  &  8.58 &  9.88&  --  & 10.58 \\
  & $n_U$      & 15.13&\pdk00& 15.16 & 16.06&\pdk10& 15.22 & 16.62&\pdk11& 16.64 & 15.34&\pdk00& 15.50 \\
  & $\omega_U$ & 41.31&  --  & 41.11 & 45.69&  --  & 41.21 & 46.07&  --  & 46.14 & 42.79&  --  & 43.21 \\

  \hline

  \multirow4*{\rep{(1,2)}}
  & $n_1$     &&& &  2.00&  --  & 13.59 &  2.00&  --  &  5.31 \\
  & $n_2$     &&& & 14.88&  --  & 13.72 &  4.51&  --  &  5.31 \\
  & $n_U$     &&& & 18.88&\pdk10& 15.15 & 16.48&\pdk11& 16.51 \\
  & $\omega_U$&&& & 28.65&  --  & 37.84 & 45.71&  --  & 45.80 \\

  \hline

  \multirow4*{\rep{(1,3)}}
  & $n_1$     &&& &  3.57&  --  & 12.99 \\
  & $n_2$     &&& & 11.01&  --  & 13.52 \\
  & $n_U$     &&& & 17.36&\pdk10& 15.18 \\
  & $\omega_U$&&& &  0.08&  --  & 32.63 \\

  \hline

  \multirow4*{\rep{(1,4)}}
  & $n_1$     &&& &  8.54&  --  & 11.23 \\
  & $n_2$     &&& & 11.38&  --  & 12.60 \\
  & $n_U$     &&& & 15.71&\pdk00& 15.30 \\
  & $\omega_U$&&& &  2.81&  --  & 18.36 \\

  \hline

  \multirow4*{\rep{(2,1)}}
  & $n_1$     &&& & 12.19&  --  & 13.23 &  2.00&  --  &  9.93 \\
  & $n_2$     &&& & 15.53&  --  & 14.29 &  9.14&  --  &  9.93 \\
  & $n_U$     &&& & 15.53&\pdk00& 15.13 & 15.81&\pdk11& 15.87 \\
  & $\omega_U$&&& & 44.11&  --  & 41.80 & 44.01&  --  & 44.17 \\

  \hline

  \multirow4*{\rep{(2,2)}}
  & $n_1$     &&& &  2.00&  --  & 12.68 &  2.00&  --  &  7.73 \\
  & $n_2$     &&& & 14.88&  --  & 13.81 &  6.42&  --  &  7.73 \\
  & $n_U$     &&& & 18.28&\pdk10& 15.15 & 15.60&\pdk01& 15.70 \\
  & $\omega_U$&&& & 28.95&  --  & 37.26 & 43.47&  --  & 43.73 \\

  \hline

  \multirow4*{\rep{(2,3)}}
  & $n_1$     &&& &  3.19&  --  & 11.61 &  2.00&  --  &  3.26 \\
  & $n_2$     &&& & 10.90&  --  & 13.15 &  2.75&  --  &  3.26 \\
  & $n_U$     &&& & 16.95&\pdk00& 15.21 & 15.32&\pdk00& 15.36 \\
  & $\omega_U$&&& &  0.09&  --  & 28.64 & 42.75&  --  & 42.85 \\

\hline\hline

\multicolumn{15}r{(Table continued to page \pageref{p:a2p3})}
\label{p:a2p2}

\end{tabular}
\end{center}

\end{table}

\begin{table}[p]
  \label{p:a2p3}

\bigskip

\footnotesize
\begin{center}
\begin{tabular}{ccc@{\hspace{1pt}}c@{\hspace{1pt}}cc@{\hspace{1pt}}c@{\hspace{1pt}}cc@{\hspace{1pt}}c@{\hspace{1pt}}cc@{\hspace{1pt}}c@{\hspace{1pt}}cc@{\hspace{1pt}}c@{\hspace{1pt}}c}

  \multicolumn7l{(\reftable{t:a2results}: continued from 
    page \pageref{p:a2p2})}  \\ 

&& \fake && \fake & \fake && \fake & \fake && \fake & \fake && \fake & \fake && \fake \\

  \hline\hline
  \multicolumn{17}c{Chains VIa to Xa} \\ 
  \hline\hline
  $(r_{10},r_{126})$
  & Quantity & \multicolumn3c{VIa} &
  \multicolumn3c{VIIa} & \multicolumn3c{VIIIa}
  & \multicolumn3c{IXa} & \multicolumn3c{Xa}  \\
  
  \hline

  \hline

  \multirow4*{\rep{(3,1)}}
  & $n_1$     &&& &&& &  2.00&  --  & 10.97 \\
  & $n_2$     &&& &&& & 10.12&  --  & 10.97 \\
  & $n_U$     &&& &&& & 15.18&\pdk00& 15.28 \\
  & $\omega_U$&&& &&& & 42.40&  --  & 42.63 \\

  \hline

  \multirow4*{\rep{(3,2)}}
  & $n_1$     &&& &  2.00&  --  & 11.65 \\
  & $n_2$     &&& & 14.88&  --  & 13.92 \\
  & $n_U$     &&& & 17.72&\pdk10& 15.16 \\
  & $\omega_U$&&& & 29.22&  --  & 36.60 \\

  \hline

  \multirow4*{\rep{(3,3)}}
  & $n_1$     &&& &  2.81&  --  &  9.91 \\
  & $n_2$     &&& & 10.80&  --  & 12.70 \\
  & $n_U$     &&& & 16.55&\pdk00& 15.25 \\
  & $\omega_U$&&& &  0.08&  --  & 23.70 \\

  \hline

  \multirow4*{\rep{(4,2)}}
  & $n_1$      &&& &  2.00&  --  & 10.47 \\
  & $n_2$      &&& & 14.88&  --  & 14.03 \\
  & $n_U$      &&& & 17.19&\pdk10& 15.16 \\
  & $\omega_U$ &&& & 29.48&  --  & 35.85 \\

  \hline

  \multirow4*{\rep{(4,3)}}
  & $n_1$     &&& &  2.53&  --  &  7.69 \\
  & $n_2$     &&& & 10.73&  --  & 12.11 \\
  & $n_U$     &&& & 16.16&\pdk00& 15.32 \\
  & $\omega_U$&&& &  0.36&  --  & 17.23 \\

  \hline

  \multirow4*{\rep{(5,1)}}
  & $n_1$     &&& &&& &&& &&& &  2.00&  --  &  4.71 \\
  & $n_2$     &&& &&& &&& &&& &  3.89&  --  &  4.71 \\
  & $n_U$     &&& &&& &&& &&& & 16.13&\pdk11& 16.03 \\
  & $\omega_U$&&& &&& &&& &&& & 38.84&  --  & 39.03 \\

  \hline

  \multirow4*{\rep{(5,2)}}
  & $n_1$     &&& &  2.00&  --  &  9.11 \\
  & $n_2$     &&& & 14.88&  --  & 14.17 \\
  & $n_U$     &&& & 16.70&\pdk10& 15.17 \\
  & $\omega_U$&&& & 29.72&  --  & 34.99 \\

  \hline

  \multirow4*{\rep{(6,2)}}
  & $n_1$     &&& &  2.00&  --  &  7.53 \\
  & $n_2$     &&& & 14.88&  --  & 14.33 \\
  & $n_U$     &&& & 16.24&\pdk10& 15.17 \\
  & $\omega_U$&&& & 29.94&  --  & 33.98 \\

\hline\hline

\end{tabular}
\end{center}

\end{table}

\begin{table}
\caption{Results for b-type symmetry breaking with two intermediate
  scales.  Notation for the separators have been explained in the
  caption of \reftable{t:a2results}.}\label{t:b2results}

\footnotesize
\begin{center}
\begin{tabular}{ccc@{\hspace{1pt}}c@{\hspace{1pt}}cc@{\hspace{1pt}}c@{\hspace{1pt}}cc@{\hspace{1pt}}c@{\hspace{1pt}}cc@{\hspace{1pt}}c@{\hspace{1pt}}cc@{\hspace{1pt}}c@{\hspace{1pt}}c}

  \hline\hline
  \multicolumn{17}c{Chains Ib to Vb} \\ 
  \hline\hline

  $(r_{10})$ 
  & Quantity & \multicolumn3c{Ib} &
  \multicolumn3c{IIb} & \multicolumn3c{IIIb}
  & \multicolumn3c{IVb} & \multicolumn3c{Vb}  \\
  
  \hline

  \multirow4*{\rep{(1)}}
  & $n_1$      & 10.41&  --  & 13.62 & 10.67&  --  & 13.71 & 10.41&  --  & 13.71 & 10.41&  --  & 13.71 & 13.05&  --  & 13.59 \\ 
  & $n_2$      & 16.72&  --  & 13.62 & 16.49&  --  & 13.71 & 16.72&  --  & 13.71 & 16.72&  --  & 13.71 & 14.20&  --  & 13.59 \\
  & $n_U$      & 16.72&\pdk10& 15.56 & 16.49&\pdk10& 15.51 & 16.72&\pdk10& 15.51 & 16.72&\pdk10& 15.51 & 15.11&\pdk00& 15.37 \\ 
  & $\omega_U$ & 46.34&  --  & 45.26 & 45.74&  --  & 44.77 & 46.34&  --  & 44.77 & 46.34&  --  & 44.77 & 44.80&  --  & 45.05 \\ 

  \hline

  \multirow4*{\rep{(2)}}
  & $n_1$      & 11.23&  --  & 13.64 & 11.41&  --  & 13.71 & 11.23&  --  & 13.71 & 11.23&  --  & 11.40 \\
  & $n_2$      & 15.97&  --  & 13.64 & 15.81&  --  & 13.71 & 15.97&  --  & 13.71 & 15.93&  --  & 11.42 \\
  & $n_U$      & 15.97&\pdk10& 15.20 & 15.82&\pdk10& 15.16 & 15.97&\pdk10& 15.16 & 15.97&\pdk11& 15.82 \\
  & $\omega_U$ & 44.42&  --  & 43.96 & 44.02&  --  & 43.58 & 44.42&  --  & 43.58 & 44.42&  --  & 44.02 \\

  \hline

  \multirow4*{\rep{(3)}}
  & $n_1$      & 11.89&  --  & 12.77 & 12.01&  --  & 12.59 & 11.89&  --  & 12.75 & 11.89&  --  & 12.00 \\
  & $n_2$      & 15.37&  --  & 14.51 & 15.27&  --  & 14.73 & 15.37&  --  & 14.59 & 15.27&  --  & 12.03 \\
  & $n_U$      & 15.37&\pdk00& 15.12 & 15.27&\pdk00& 15.12 & 15.37&\pdk00& 15.12 & 15.37&\pdk00& 15.27 \\
  & $\omega_U$ & 42.89&  --  & 42.83 & 42.62&  --  & 42.57 & 42.89&  --  & 42.70 & 42.88&  --  & 42.62 \\
  \hline\hline

  \\
  \\

  \hline\hline
  \multicolumn{17}c{Chains VIb to Xb} \\ 
  \hline\hline

  $(r_{10})$ 
  & Quantity & \multicolumn3c{VIb} &
  \multicolumn3c{VIIb} & \multicolumn3c{VIIIb}
  & \multicolumn3c{IXb} & \multicolumn3c{Xb}  \\
  
  \hline

  \multirow4*{\rep{(1)}} 
  & $n_1$      & 13.24&  --  & 13.71 & 13.63&  --  & 13.71 &  2.00&  --  & 10.11 &  2.00&  --  & 11.07 &  2.00&  --  & 12.45 \\
  & $n_2$      & 14.14&  --  & 13.71 & 15.32&  --  & 13.78 &  9.92&  --  & 10.11 & 10.89&  --  & 11.07 & 12.29&  --  & 12.45 \\
  & $n_U$      & 15.11&\pdk00& 15.28 & 15.39&\pdk00& 15.35 & 16.70&\pdk11& 16.71 & 16.09&\pdk11& 16.12 & 15.67&\pdk11& 15.72 \\
  & $\omega_U$ & 44.45&  --  & 44.39 & 45.05&  --  & 44.65 & 46.29&  --  & 46.31 & 44.72&  --  & 44.79 & 45.30&  --  & 45.36 \\

  \hline

  \multirow4*{\rep{(2)}} 
  & $n_1$     &&& &&& &  2.00&  --  & 11.01 &  2.00&  --  & 11.68 &  2.00&  --  & 12.71 \\
  & $n_2$     &&& &&& & 10.81&  --  & 11.01 & 11.49&  --  & 11.68 & 12.55&  --  & 12.71 \\
  & $n_U$     &&& &&& & 15.94&\pdk11& 15.96 & 15.53&\pdk00& 15.56 & 15.25&\pdk00& 15.29 \\
  & $\omega_U$&&& &&& & 44.34&  --  & 44.38 & 43.28&  --  & 43.37 & 43.87&  --  & 43.94 \\

  \hline

  \multirow4*{\rep{(3)}} 
  & $n_1$     &&& &&& &  2.00&  --  & 11.73 \\
  & $n_2$     &&& &&& & 11.52&  --  & 11.73 \\
  & $n_U$     &&& &&& & 15.34&\pdk00& 15.36 \\
  & $\omega_U$&&& &&& & 42.79&  --  & 42.84 \\

\hline\hline

\end{tabular}
\end{center}

\end{table}


\subsection{Two intermediate scales}
We now turn to the symmetry breaking chains which involve two
intermediate stages.  As is well known, it is not possible to uniquely
solve the values of the various breaking scales in this case.  Since
there are only three boundary conditions corresponding to the
experimentally measured values of the SM gauge couplings at the weak
scale, while there are four unknown variables, namely $n_1$, $n_2$,
$n_U$ and $\omega_U$, one can at most find some regions allowed by the
selection criteria for each of them.  All solutions which pass these
tests have been listed.  \reftable{t:a2results} contain results for
a-type breaking chains, whereas \reftable{t:b2results} contain results
for b-type breaking chains.  Since the number of possible chains is
huge, we had to break each of these tables into two parts for
convenience of display, first giving results for chains I to V, and
next for chains VI to X.  For chains XI and XII, we find no solution,
so they do not appear in the tables.

In the results presented in \reftable{t:a2results} and
\reftable{t:b2results}, different quantities given on the left sides
of the dashes as a set, correspond to one solution, whereas those
of the right sides correspond to another solution.  Thus, for example,
with chain Ia and $(r_{10},r_{126})=(\rep{1,1})$, one extreme of the
solution range lies at $n_1=9.02$ and $n_2=n_U=16.66$, whereas the
other extreme solution is at $n_1=11.17$, $n_2=11.19$ and $n_U=16.75$.
Of course, all intermediate values are allowed.  Since the RG
equations are linear in the scales and also in $\omega_U$, the values
of $n_2$, $n_U$ and $\omega_U$ corresponding to any intermediate value
of $n_1$ can be obtained by linear interpolation.

As for the cases with single intermediate stage, the solutions with
$r_{10}=r_{126}=\rep1$ merely reproduce the solutions obtained in the
absence of $H$.  Our results agree roughly with the results in earlier
1-loop calculations \cite{Deshpande:1992em}, except for chain XIIa
which is ruled out anyway. Slight deviations are due to reasons
mentioned in the context of symmetry breaking with one intermediate
scale.  Also, in chains VIII to XII where there are two U(1) factors,
we take kinetic mixing into account.

\section{Discussions}\label{s:discu}
Here we briefly outline some important aspects and limitations of this
work.
\begin{enumerate}
\item It has to be understood that our results are based on 1-loop
  calculations only. It is known that higher-loop calculation and
  threshold corrections may sometimes produce large changes in the RG
  solutions \cite{Bertolini:2009qj, Meloni:2019jcf}. To leave room for
  the effect of 2-loop corrections we somewhat relaxed the proton
  decay bound using an order of magnitude estimation.  We emphasize
  that 2-loop estimation and 1-loop threshold corrections carry
  additional baggage of uncertainties due to the involvement of more
  parameters arising from specific details of model spectra. All these
  effectively relax the proton decay bound, which constitutes the
  basis of our strategy.

\item   Since the Higgs fields are charged under $H$, their
  VEVs indeed break $H$. In fact, $H$ might have already been broken
  by the VEV of an SO(10) singlet at an even higher scale. The question
  is whether all Higgs masses are pushed up to that scale by a na\"ive
  application of the extended survival hypothesis. We must remember
  that this hypothesis is at best a guess that
  one applies in the absence of knowledge of the detailed Higgs
  potential guided by symmetries. In this paper, whenever we refer to
  some $H$, we assume that multiplicities, arising due to $H$, of a
  given $n$-plet of SO(10) are treated identically in the
  potential. Then, since a $10$-plet is lighter than the $126$-plet as
  a consequence of SO(10) breaking sequence, all copies $10$-plets are
  lighter than all copies of $126$-plets. This is our extended
  survival hypothesis for the present analysis. Note, we remained
  agnostic about $H$ throughout the paper. We neither mentioned if $H$
  breaks to nothing or to some subgroup, nor did we attempt to write
  down a full potential involving $H$ as a symmetry. The latter would
  have been horrifically complicated and severely model
  dependent. Even if one is not ready to accept our hypothesis, our
  analysis still goes through, as our primary objective is to examine
  how many light copies of SO(10) $n$-plets can be kicked into life to
  contribute to RG evolution at different stages, regardless of what
  $H$ is. The extreme opposite scenario that all copies, except a lone
  generation of SO(10), become super-heavy at the topmost scale where
  $H$ breaks, however, deprecates any multiplicity of states, as if
  there is no horizontal symmetry, thus trivializing our
  analysis. However, this is only a limiting scenario.

\item From our 1-loop results, one trend is quite clear: too many
  copies of light Higgs multiplets is not good for a model.  For
  a-type symmetry breaking, employing 126-plets leading to
  renormalizable Yukawa interactions, more than three copies of the
  10-plet or the 126-plet works only for very few chains. For b-type
  symmetry breaking, which utilizes 16-plets of Higgs bosons instead
  of 126, we obtain results for at most three copies of the 10-plet.

\item In some of the symmetry breaking chains, the $U(1)_{B-L}$ factor
  appears explicitly in the intermediate stages.  In some others, it
  is part of an SU(4) factor.  For the first set of chains, our
  results show that, if $r_{10}=r_{126}=\rep1$, most of these chains
  yield a large value of $M_1\sim10^9$ GeV, except chains VIIIa and
  IXa predicting much lower $M_1$ around TeV scale. However, for
  $r_{126} > \rep1$, from \reftable{t:a1results} and
  \reftable{t:a2results} we observe that the $\rm U(1)_{B-L}$ breaking
  scale $M_1$ can be brought to much lower values, within the reach of
  collider experiments. The presence of a TeV scale Abelian gauge
  boson ($Z'$) in those cases may have testable signatures in various
  flavor and electroweak observables. A detailed discussion is beyond
  the mandate of this paper.

\item To make any speculation and draw conclusion on $H$ we need to
  make further assumption that the splitting pattern among Higgs
  masses is controlled purely by the pattern of SO(10) breaking
  sequence. For example, consider SO(3) as the horizontal symmetry
  group. For SO(3), the symmetric combination of $\rep{3\times 3}$
  contains a \rep5-plet and a singlet. We can conclude that SO(3) is
  disfavored if all the \rep5 copies of the 10-plet remain light till
  $M_Z$, since \reftable{t:a2results} shows $r_{10}=\rep{5}$
  necessarily requires $r_{126}=\rep{2}$. The other escape route is to
  consider two copies of 126-dimensional Higgs which are singlet under
  SO(3). If, however, the potential is such that some horizontal
  copies receive much higher masses, thereby reducing the multiplicity
  contributing to RG in intermediate stages, then our previous
  observation of SO(3) being disfavored gets accordingly
  diluted. Analogous conclusions can also be drawn for other choices
  of $H$, e.g. SU(3).
 
\item It is of course not necessary to think in terms of irreps only.
  If, for example, we need 4 copies of the 10, they need not form a
  4-dimensional irrep of $H$.  One can have a triplet and a singlet,
  or maybe two doublets.  All we need is that they are part of the
  symmetric product of two triplets of $H$, since we are dealing with
  three generations.  Alternatively, the fermions can also live in
  reducible representations under $H$. For example, if the horizontal
  symmetry is $S_3$, the SM quarks and leptons can transform as
  $\rep{2+1}$, where the third generation fermions form the
  singlet. This might be useful in order to fulfill the
  phenomenological requirement of reproducing the correct flavor
  hierarchy.

  Additional SO(10) Higgs multiplets, other than 10- and
  126-plets, might also be advocated to meet some phenomenological
  criteria in the context of flavor physics, which can affect the
  unification of the gauge couplings. Here, we consider the minimal
  possibility from the perspective of grand unification only.

\item A few comments on the issue of predictability of these models
  are in order. In the absence of any horizontal symmetry, any SO(10)
  predictions relating fermion masses must be confined within the same
  family, because the couplings of different generations would be
  independent. Horizontal symmetry provides an option for interfamily
  predictions. Note that if we put all the three families in a single
  irrep, the number of Yukawa couplings reduces dramatically rendering
  the model superpredictable. This is too restrictive and already
  falsified by data.  On the contrary, if we put the fermion and/or
  scalar families in reducible representations having multiple irreps,
  as stressed earlier, the tension eases a bit. For example, the
  triplet \rep3 representation of the discrete group $S_3$ is a
  reducible representation: \rep{3 = 2 + 1}. If we put the first two
  fermion families in the doublet of $S_3$, and the third family as
  the singlet, the observed heaviness of the third generation can be
  accounted for, while leaving the splitting between the first two
  families to additional model building.  In other words, while the
  intention behind imposing horizontal symmetry is to reduce the
  number of parameters to increase predictability, real world data
  compel us to take shelter under the cover of multiple irreps at the
  expense of introducing a few more parameters. Though we lose some
  predictability this way, nevertheless, the number of such parameters
  never exceeds or even touches the corresponding number for
  conventional three-family SO(10) without any family symmetry at
  all. It is this issue that underscores the need for a horizontal
  symmetry.

\item It is not just a question of fermion masses.  There are other
  issues in flavor physics, like fermion mixing and CP violation.
  For them as well, multiple irreps might prove essential.  For
  example, in the context of nonunified models multiple Higgs
  doublets motivated by some horizontal symmetry, have been employed
  in the literature to deduce the general nature of the CKM matrix
  \cite{Canales:2012ix, Canales:2013cga, Das:2015sca,
    Cogollo:2016dsd}.  Various models predicting the masses and
  mixing in the neutrino sector also utilize more than one Higgs
  multiplets with varying degree of success. Similarly in the GUT
  models, existence of reducible representations of $H$ would
  definitely be an added advantage to predict the observed flavor
  texture. Moreover an enriched scalar sector may predict novel
  phenomenological signatures, which are otherwise unavailable. For
  example, more copies can give additional freedom of accommodating CP
  violation at low scale.

\item Values of $r_{10}$ larger than \rep{1} imply multiple Higgs
  doublets with masses around the weak scale. Such scenarios may
  receive severe constraints from flavor changing neutral current
  processes, from the direct searches of charged Higgs bosons, etc.
  Thus our selection criteria can, in principle, confront the presence
  of multi-Higgs models at the weak scale.

\item In this paper we work with three generations of chiral fermion,
  though our analysis can be easily extended to accommodate more
  generations.  We have observed that for four chiral generations,
  charged under $H$, even fewer chains of symmetry breaking would be
  allowed.

\item As a final comment, we mention that our analysis can also be
  extended to GUT models based on $E_6$ and larger groups  as well
  \cite{Camargo-Molina:2016bwm,Camargo-Molina:2016yqm,Camargo-Molina:2017kxd,Morais:2020ypd,Morais:2020odg},
  which might constitute an interesting exercise.

\end{enumerate}

\subsection*{Acknowledgments}
A.B. acknowledges discussions with Triparno Bandyopadhyay, and support
from the Department of Atomic Energy, Government of India.  G.B.
acknowledges support of the J.C. Bose National Fellowship from the
Department of Science and Technology, Government of India (SERB Grant
No. SB/S2/JCB-062/2016).  The work of P.B.P. was supported by the SERB
Grant EMR/2017/001434 of the Government of India.


\bibliographystyle{JHEP}
\bibliography{so10gen}

\end{document}